 \documentclass[smallabstract,smallcaptions,12pt]{dccpaper}
\usepackage{epsfig}
\usepackage{citesort}
\usepackage{amsmath}
\usepackage{amssymb}
\usepackage{color}
\usepackage{url}
\usepackage{multirow}

\newlength{\figurewidth}
\newlength{\smallfigurewidth}
\graphicspath{{Figures/}}
\newcommand{\eqn}[1]
{(\ref{#1})}

\title{\large\textbf{Orthonormal Matrix Codebook Design for Adaptive Transform Coding}}
\author{%
Rashmi Boragolla, and Pradeepa Yahampath\\[0.5em]
{\small\begin{minipage}{\linewidth}\begin{center}
\begin{tabular}{c}
Department of Electrical and Computer Engineering, University of Manitoba \\
75 Chancellors Circle, Winnipeg MB R3T 2N2, Canada \\
\url{borbnwmr@myumanitoba.ca}, \url{pradeepa.yahampath@umanitoba.ca} 
\end{tabular}
\end{center}\end{minipage}}
}

\begin{document}
\maketitle

\begin{abstract}
A novel algorithm for designing optimized orthonormal transform-matrix codebooks for adaptive transform coding of a non-stationary vector process is proposed. This algorithm relies on a block-wise stationary model of a non-stationary process and finds a codebook of transform-matrices by minimizing the end-to-end mean square error of transform coding averaged over the distribution of stationary blocks of vectors. The algorithm, which belongs to the class of block-coordinate descent algorithms, solves an intermediate minimization problem involving matrix-orthonormality constraints in a computationally efficient manner by mapping the problem from the Euclidean space to the Stiefel manifold. As such, the algorithm can be broadly applied to any adaptive transform coding problem. Preliminary results obtained with inter-prediction residuals in an H265 video codec are presented to demonstrate the advantage of optimized adaptive transform codes over non-adaptive codes based on the standard DCT.
\end{abstract}

\section{Introduction}

\label{sec:intro}
It is well-known that the optimal transform for coding stationary Gaussian sources based on the mean squared error (MSE) is the Karhunen-Lo\'{e}ve transform (KLT) \cite{gersho1991vector,goyal2000transform}. However, in real-world applications where the source data is highly non-stationary, the tendency has been to use generic fixed-transforms such as the discrete cosine transform (DCT). Nonetheless, adaptive transform coding (ATC) remains an active research area, and recently, some forms of ATC have also been incorporated into newer video coding standards such as H265. ATC methods reported in the literature are very diverse and typically specialized to signals with certain characteristics, such as images containing various directional properties \cite{Chang,Zeng}, and video residuals produced by intra-prediction \cite{Ye,Han,Zhao,Zou} or motion-compensation (MC) \cite{Biatek,Wang,Acao2014singular,gu2012rotated,kamisli20101,Lan}. Others have considered the design of more general transforms using sparsity constraints \cite{sezer2015approximation,Kang}. However, such approaches do not take into account the effect of quantization errors due to finite rate compression.

In contrast to previous work, we present a novel algorithm for designing a fixed-size {\em codebook} of orthogonal transform matrices based on a fairly general block-wise stationary model of a non-stationary vector process that exhibits local stationarity properties. In this case, the best transform matrix from the codebook is adaptively chosen for each stationary block. The matrix codebook is designed to minimize the block-MSE, averaged over an ensemble of blocks representing the non-stationary process. The codebook optimization is carried out by applying the algorithm to a training set of blocks. The only requirement to use this algorithm is the availability of a differentiable expression for the quantization MSE of the transform coefficients expressed in terms of the transform matrix. To this end, we make the general assumption that the probability distribution of the transform coefficients is solely characterized by the mean and the variance (e.g., Gaussian and Laplace distributions). While we have focused on uniform quantization which has been widely used in practice, non-uniform quantization may also be incorporated. So far, we have only considered the design of non-separable transforms.

The proposed algorithm belongs to the class of {\em  block coordinate-descent} (BCD) algorithms \cite{Bertsekas}. In contrast to other approaches previously considered to solve related matrix-codebook design problems \cite{dony1995optimally,caglar1998vq,effros1999weighted,archer2004generalized}, in our formulation we map the harder problem of minimizing an objective $\Theta(\pmb{T})$ subject to an orthonormality constraint on the matrix argument $\pmb{T}$, from the Euclidian space to a simpler unconstrained problem on the {\em Stiefel manifold} \cite{Manton}. This approach, which explicitly minimizes the transform coding MSE with respect to the matrix codebook, is superior to other indirect methods used to enforce orthonormality constraints, such as the Givens parameterization where a set of rotation angles rather than orthogonal matrices have to be quantized \cite{Sadrabadi}. Being a BCD algorithm based on the exact matrix-derivative of the objective function, our algorithm is guaranteed to converge to a (locally) optimal transform-matrix codebook.

We have so far tested the proposed algorithm on MC residuals obtained from an H265 video codec, which contains highly non-stationary segments. Results obtained with a substantial test set of videos showed that ATC based on proposed codebook designs consistently outperformed non-ATC based on the standard DCT. A selected set of results are presented in Section \ref{sec: exp_results}. 

\section{Adaptive Transform Coding Setup}
\label{sec:TCnATC}


First, consider the standard transform coding problem, as applied to a stationary random vector $\pmb{X} \in \mathbb{R}^k$. In the encoding process, rather than directly quantizing $\pmb{X}$, it is first linearly transformed to obtain $\pmb{Y} = \pmb{T}\pmb{X}$, and the transform coefficients $\pmb{Y}=(Y_1,\ldots,Y_k)^T$ are then scalar quantized and entropy coded, where $\pmb{T} \in \mathbb{R}^{k \times k}$ is the orthonormal transform matrix. In this work, we assume that uniform scalar quantizers are used. Let the step-size of the quantizer for the transform coefficient $Y_i$ be $\Delta_i, i = 1, ..., k$ and denote the quantized version of $\pmb{Y}$ by $\hat{\pmb{Y}}=(\hat{Y}_1,\ldots,\hat{Y}_k)^T$. The reconstructed source vector is thus $\hat{\pmb{X}} = \pmb{T}^{-1}\hat{\pmb{Y}}$. Assume that each transform coefficient is zero mean and that the quantization MSE is a function of the quantizer's input variance $\sigma_{Y_i}^2=E[ Y_i^2 ]$. Accordingly, denote the quantization MSE of the coefficient $Y_i$ by $\theta(\sigma_{Y_i}^2,\Delta_i)=E(Y_i -\hat{Y}_i)^2$. Let the variances of the transform coefficients be $(\sigma_{Y_1}^2,\ldots,\sigma_{Y_k}^2)^T$. These variances are the diagonal elements of $\pmb{C}_{\pmb{Y}}=\pmb{T}\pmb{C}_{\pmb{X}}\pmb{T}^T$, where $\pmb{C}_{\pmb{X}}=E [ \pmb{X}\pmb{X}^T ]$ is the covariance matrix of $\pmb{X}$. The MSE of transform coding $\pmb{X}$ is given by $\Theta(\pmb{T},\pmb{C}_{\pmb X})=E \Vert \pmb{X}-\hat{\pmb{X}} \Vert^2=\sum_{i=1}^k\theta(\sigma_{Y_i}^2,\Delta_i)$, where we make it explicit the dependence of $\Theta$ on $\pmb{C}_{\pmb X}$ through the variances $\sigma_{Y_i}^2$. We will assume that $\Delta_i$, $i=1,\ldots,k$ are given. In this case, the optimal $\pmb{T}$ can be found by solving

\begin{align}
  & \underset{\pmb{T}}{\mbox{Minimize }} \Theta(\pmb{T},\pmb{C}_{\pmb{X}}) \label{optim_tr}\\
 &  \mbox{subject to } \pmb{T}^T\pmb{T}  =\pmb{I}_k \mbox{ (orthonormality constraint)}, \nonumber
\end{align}
where $\pmb{I}_k$ is the $k\times k$ identity matrix. It is easy to demonstrate that if $\pmb{X}$ is Gaussian, regardless of $\Delta$, the KLT is the solution to \eqn{optim_tr} \cite{gersho1991vector,goyal2000transform}.

Next, consider coding a sequence of non-stationary random vectors. In this case, the optimal transform may vary with the input vector, and adaptive coding can significantly improve the rate-distortion performance. We model the non-stationary process by a {\em block-wise} stationary vector process and encode each stationary block of vectors using a single transform optimized to local statistics. To state the problem more concretely, let $B \in \mathbb{S}_{B}$ be a locally stationary block of vectors in the non-stationary process, where $\mathbb{S}_{B}$ is the set of all possible (ensemble of) blocks. Also, let ${\pmb C}_B$ be the covariance matrix for $\pmb{X} \in B$, and ${\mathbb S}_{\pmb C}$ be the ensemble of covariance matrices. Obviously, we can determine the optimal transform matrix for each ${\pmb C}_B$ in the ensemble by solving \eqn{optim_tr} on a per-block basis. For example, if $\pmb{X}$ is a Gaussian vector, then the optimal transform for stationary block $B$ is the KLT of ${\pmb C}_B$ \cite{gersho1991vector,goyal2000transform}. Instead, we use a codebook of orthonormal transform matrices $\mathcal{T} = \{\tilde{\pmb{T}}_1, \tilde{\pmb{T}}_2,...,\tilde{\pmb{T}}_N \}$ and the best transform matrix for a given block $B$ is selected from the codebook as
 $ \pmb{T}^* = \arg \min_{\pmb{T} \in \mathcal{T} } \frac{1}{|B|}\sum_{\pmb X \in B} \Vert \pmb{X}-\hat{\pmb X}(\pmb{T}) \Vert^2$ 
where $\hat{\pmb X}(\pmb{T})$ is the reconstructed version of source vector $\pmb{X}$ using the transform $\pmb{T}$ and $\vert B \vert$ is the number of vectors in $B$.

\section{Transform Matrix Codebook Optimization}\label{sec_tr_optim}
Let us consider the problem of designing the optimal codebook ${\mathcal T}$ for a given block ensemble ${\mathbb S}_B$. We note that the design problem essentially involves partitioning ${\mathbb S}_B$  into $N$ non-overlapping subsets and assigning a single transform matrix to each subset, such that the MSE of transform coding, averaged over the ensemble is minimized. As emphasized in \eqn{optim_tr}, the MSE of block $B$ is a function of the covariance matrix $\pmb{C}_B$. Thus, an equivalent codebook design problem is to partition  the covariance matrix ensemble ${\mathbb S}_C$ into $N$ non-overlapping subsets ${\mathcal G}=\{ \Omega_1,\ldots,\Omega_N\}$, where $\Omega_i \subset {\mathbb S}_C$, and to determine the optimal codeword $\tilde{\pmb{T}}_i$ for each $\Omega_i$, $i=1,\ldots,N$. Assume that $\pmb{C}_{B}$ is distributed over ${\mathbb S}_C$ according to some probability density function. Then, the MSE of encoding all stationary blocks $B \in {\mathbb S}_B$ using the transform codebook ${\mathcal T}$ is
\begin{align}
  \bar{\Theta}(\mathcal{T},{\mathcal G}) &= \sum_{i=1}^N E \left[ \Theta(\tilde{\pmb{T}}_i, \pmb{C}_B)| \pmb{C}_B \in \Omega_i \right] P(\Omega_i), \label{eqn_MSE_ATC1}
\end{align}
where the expectation is taken over the distribution of $\pmb{C}_B$. We wish to find the codebook $\mathcal{T}^*$ which minimizes \eqn{eqn_MSE_ATC1} with respect to the sets $(\mathcal{T},{\mathcal G})$. Specifically, we seek to solve
\begin{align}
  \underset{\mathcal{T},{\mathcal G}}{\mbox{Minimize }} & \bar{\Theta}(\mathcal{T},{\mathcal G})\label{eqn_TCB}\\
 \mbox{ subject to } & \tilde{\pmb{T}}_i^T\tilde{\pmb{T}}_i=\pmb{I}_k, \ i=1,\ldots,N. \nonumber 
\end{align}
The direct solution of this constrained minimization problem appears intractable. The problem is reminiscent of codebook design in vector quantization, where a BCD algorithm, commonly referred to as the generalized Lloyd's algorithm, is used to solve a similar problem \cite{gersho1991vector}. However, in that case there are no constraints on code vectors. In this paper, we present a BCD algorithm incorporating orthogonal constraints on the matrix codebook to solve \eqn{eqn_TCB}, where one alternates between the solutions to two sub-problems described below, such that the algorithm converges to a solution of \eqn{eqn_TCB}.

\subsection{Sub-problem 1: Optimal partition ${\mathcal G}$ for a fixed codebook ${\mathcal T}$}
It is straightforward to argue that, for a fixed ${\mathcal T}$, the optimal ${\mathcal G}$ that minimizes $\bar{\Theta}(\mathcal{T},{\mathcal G})$ is given by ${\mathcal G}^*=\{ \Omega^*_1,\ldots,\Omega^*_N\}$, where
\begin{align}
  \Omega^*_i = \{ \pmb{C}_B \in \mathbb{S}_{\pmb{C}} : \Theta(\tilde{\pmb{T}}_i,\pmb{C}_B)< \Theta(\tilde{\pmb{T}}_j,\pmb{C}_B) \ \forall j \neq i \} \label{partition}
\end{align}
with ties broken suitably. Note that the orthogonality constraint is irrelevant to this sub-problem.

\subsection{Sub-problem 2: Optimal codebook ${\mathcal T}$ for a fixed partition ${\mathcal G}$}
Given the partition set ${\mathcal G}$, $\tilde{\pmb{T}}_i$ only affects the $i$-the term of the sum in \eqn{eqn_MSE_ATC1}. Thus, the codebook ${\mathcal T}^*= \{\tilde{\pmb{T}}^*_1, \tilde{\pmb{T}}^*_2,...,\tilde{\pmb{T}}^*_N \}$ that minimizes $\bar{\Theta}(\mathcal{T},{\mathcal G})$  is given by
\begin{align}
  \tilde{\pmb{T}}^*_i=\arg \underset{\pmb{T} \in {\mathbb R}^{k \times k}}{\min} E \left[ \Theta(\pmb{T}, \pmb{C}_B)| \pmb{C}_B \in \Omega_i \right] \mbox{ subject to } \pmb{T}^T\pmb{T}=\pmb{I}_k, \label{cond_centroid}
\end{align}
$i=1,\ldots,k$. While the solution to this constrained minimization problem does not appear to be straightforward, we note that the solution space, the set of all $k\times k$ orthogonal matrices, is the compact manifold referred to as the {\em orthogonal group} ${O}(k)$ \cite{Manton}. Therefore, rather than solving \eqn{cond_centroid} as a constrained minimization problem on the Euclidian space ${\mathbb R}^{k \times k}$, we can equivalently solve it as an unconstrained minimization problem on ${O}(k)$. Optimization on manifolds is a widely studied problem, see \cite{Manton} and references therein. We have used a low-complexity, modified steepest descent algorithm on $O(k)$ \cite{manton2002optimization} to solve \eqn{cond_centroid}. This algorithm requires that the objective function in \eqn{cond_centroid} be differentiable. That is, $\Theta(\pmb{T}, \pmb{C}_B)$, which is the quantization MSE associated with coding a vector with a covariance matrix $\pmb{C}_B$ using the transform matrix $\pmb{T}$, must be differentiable with respect to $\pmb{T} \in {\mathbb R}^{k\times k}$. We identify in the following two models for $\Theta(\pmb{T}, \pmb{C}_B)$ satisfying this requirement, both analytically simple enough to be incorporated in to the manifold steepest-descent algorithm.

First, let  $\pmb{Y}({\pmb{T}})={\pmb{T}}\pmb{X}$ be the coefficient vector obtained by transforming  $\pmb{X} \in B$ using the transform matrix ${\pmb{T}}$, and let the variances of these transform coefficients be $\sigma_{Y_{j}}^2({\pmb T})$, $j=1,\ldots,k$, which are the diagonal elements of the matrix ${\pmb{T}}\pmb{C}_{B}{\pmb{T}}^T$.
\vspace{1ex}

\noindent 1) {\em High-rate Gaussian model}: Given a target rate (quantizer output entropy) $R_0$ bits/vector, the minimum MSE (MMSE) of transform coding a Gaussian vector is \cite{gersho1991vector,goyal2000transform}
\begin{align}
  \Theta({\pmb{T}},\pmb{C}_B) &=\frac{k\pi e}{6}\left( \prod_{j=1}^k \sigma^2_{Y_{j}}({\pmb{T}})\right)^{1/k} 2^{-\frac{2R_0}{k}}. \label{high_rate_model} 
\end{align}
Since $R_0$ is only a scaling factor, it can be ignored. In this case, we can show that
\begin{align*}
  D_{\pmb T} \left(E[\Theta(\pmb{T}, \pmb{C}_B)]\right) &= E\left[\sum_{i=1}^k 2\pmb{e}_i \pmb{e}_i^T \pmb{T}\pmb{C}_{B} \left( \prod_{j=1,j\neq i}^k \pmb{e}_j^T \pmb{T} \pmb{C}_{B} \pmb{T}^T \pmb{e}_j\right)\right],
\end{align*}
where $D_{\pmb T}\left(f({\pmb T})\right) \in {\mathbb R}^{k \times k}$ denotes the matrix of derivatives of function $f: {\mathbb R}^{k \times k} \to {\mathbb R}$ with respect to the elements of $\pmb{T} \in {\mathbb R}^{k \times k}$, see \cite[Eqn. (2)]{manton2002optimization} for the definition, and  $\pmb{e}_i$ is the $i^{th}$ column of $\pmb{I}_k$.
\vspace{1ex}

\noindent 2) {\em Laplacian model}: An alternative unimodal pdf often used as a model for image transform coefficients is the Laplace pdf. We can show that, the MSE of quantizing a mean-zero Laplace variable with variance $\sigma^2$ using a uniform quantizer with a dead zone $(-\frac{z}{2},\frac{z}{2})$ and quantization step-size $\Delta$ is
\begin{align*}
  \theta(\sigma^2,\Delta,z) &= 2b^2 - e^{-z/2b} \left( \dfrac{z^2-\Delta^2}{4} +zb+\Delta b \dfrac{(e^{\Delta/b}+1)}{(e^{\Delta/b}-1)}\right),
\end{align*}
where $b= \sqrt{\frac{\sigma^2}{2}}$, and we have assumed that the quantizer has infinite-support. This assumption is reasonable when variable rate coding is used. In this case, we have
\begin{align}
  \Theta({\pmb{T}},\pmb{C}_B) &= \sum_{j=1}^k \theta(\sigma_{Y_{j}}^2,\Delta_j,z_j), \label{Laplacian_model}
\end{align}
and it can be shown that 
\begin{align*}
  D_{\pmb T} \left( E[\Theta(\pmb{T}, \pmb{C}_B)] \right) &= E\bigg[\sum_{j=1}^k \bigg\{ 4b_j - \dfrac{e^{-z_j/2b_j}z_j}{2b_j^2}\bigg( \dfrac{z_j^2-\Delta_j^2}{4} +z_j b_j +\Delta_j b_j \dfrac{(e^{\Delta_j/b_j}+1)}{(e^{\Delta_j/b_j}-1)}\bigg) 
    \\ & \:\:- e^{-z_j/2b_j} \bigg( z_j + \dfrac{\Delta_j b_j e^{2\Delta_j/b_j} - \Delta_j b_j + 2\Delta_j^2e^{\Delta_j/b_j}}{b(e^{\Delta_j/b_j}-1)^2} \bigg)\bigg\} \dfrac{\pmb{e}_j \pmb{e}_j^T \pmb{T}\pmb{C}_{B}}{2b_j}\bigg].
\end{align*}

\noindent {\em Steepest-descent minimization of \eqn{cond_centroid} on $O(k)$}\\
The constrained minimization problem \eqn{cond_centroid} can be equivalently stated as an unconstrained minimization problem on $O(k)=\{ \pmb{T}\in {\mathbb R}^{k\times k} : {\pmb T}^T\pmb{T} =\pmb{I}_{k} \}$ which is the set of all $k\times k$ real orthogonal matrices. Specifically,
\begin{align}
  \tilde{\pmb{T}}^*_i=\arg \underset{\pmb{T} \in O(k)}{\min} E \left[ \Theta(\pmb{T}, \pmb{C}_B)| \pmb{C}_B \in \Omega_i \right]. \label{cond_centroid2}
\end{align}
The modified steepest-descent algorithm \cite[Algorithm 15]{manton2002optimization} can be used to minimize a differentiable function on the complex Stiefel manifold $St(k,n)=\{ \pmb{T} \in {\mathbb C}^{k \times n}: \pmb{T}^H\pmb{T}=\pmb{I}_{k}\}$, the set of all $k \times n$ complex matrices whose columns are orthonormal vectors. This algorithm, which uses the Armijo's step-size rule, almost always converges to a local minimum.  As $O(k)$ is a special case of $St(k,n)$, we adapted \cite[Algorithm 15]{manton2002optimization} to solve \eqn{cond_centroid2}. Due to space limitations, we will not elaborate on specifics here, but refer the reader to \cite{manton2002optimization}.

\subsection{Complete CD Algorithm for Transform Codebook Design}
\noindent{\em Given}: A training set of covariance matrices $\tilde{S}$ representing ${\mathbb S}_C$, an initial codebook of orthonormal matrices $\mathcal{T}^{(0)} = \{\tilde{\pmb{T}}_1^{(0)},...,\tilde{\pmb{T}}_N^{(0)} \}$, a tolerance parameter $\epsilon>0$, and maximum allowed iterations $M$.  Set iteration index $t=1$.
\begin{enumerate}
\item Given $\mathcal{T}^{(t-1)}$, partition $\tilde{S}$ into $N$ subsets $\{\Omega^{(t)}_1,\ldots, \Omega^{(t)}_N\}$ according to \eqn{partition}.
\item Given $\{\Omega^{(t)}_1,\ldots, \Omega^{(t)}_N\}$, find the optimal transform codebook $\mathcal{T}^{(t)}$ by solving \eqn{cond_centroid2}.
\item Estimate (by sample averaging) $\bar{\Theta}^{(t)}$, see \eqn{eqn_MSE_ATC1}. If $\dfrac{\bar{\Theta}^{(t-1)}-\bar{\Theta}^{(t)}}{\bar{\Theta}^{(t-1)}} \leqslant \epsilon$ or $t > M$ stop; Otherwise, let $ t \to t+1$ and  repeat from 1.
\end{enumerate}

\section{Experimental Results}
\label{sec: exp_results}
\subsection{A toy example}
While the KLT is the optimal transform for coding Gaussian vectors, it is known that the KLT is not optimal for vectors from a Gaussian mixture \cite{Effros2}. Using the proposed algorithm, we optimized a single transform matrix for 2-dimensional vectors drawn from a Gaussian mixture with 3 mean-zero components  whose covariance matrices are
\begin{align*}
\pmb{C}_1 = \begin{bmatrix}
1.54  & -1.84 \\
-1.84  &  2.62 
\end{bmatrix} , \
\pmb{C}_2 = \begin{bmatrix}     
0.46  &  0.40 \\
0.40  &  0.70
\end{bmatrix} , \mbox{ and } 
\pmb{C}_3 = \begin{bmatrix}
2.22  &  0.77 \\
0.77  &  0.38  
\end{bmatrix}.  
\end{align*}
 Tables \ref{tab:Toy_ex} compares the SNR and the entropy estimated from a test set of $3 \times 10^6$ vectors. The SNR is defined as $10\log_{10}\frac{\sum \Vert \pmb{X} \Vert^2}{\sum\Vert \pmb{X}-\hat{\pmb{X}}\Vert^2}$ with sums taken over the test set. The step-size $\Delta$ was chosen such that the theoretically expected rate is 0.6 bits/sample. {\em High-rate} and {\em Laplace} respectively refer to transforms optimized using the high-rate Gaussian model \eqn{high_rate_model} and the Laplacian model \eqn{Laplacian_model}. Note that the KLT has been computed from a single covariance matrix, estimated from the complete dataset. While the result itself is not surprising, our design algorithm does find a transform matrix noticeably better than the KLT.

\subsection{Preliminary Results for Motion-compensated (MC) Video Residuals}
This section presents experimental results obtained with MC (inter-frame prediction) residuals of video sequences. Since our transform codebook design algorithm in its current form is applicable only to non-separable transforms, we considered transform coding of $4\times 4$ pixel blocks in our experiments to keep the computational complexity low. Thus we would encode 16-dimensional vectors using a $16\times 16$ transform matrix. For the purpose of generating residual frames to be used in the training of transform codebook design, we used the HM test model, \cite{HMmodel} and a set of $9$ standard CIF resolution ($352 \times 288$), 30 fps gray-scale video sequences ({\em Bus, Coastguard, Crew, Football, Foreman, Mobile, Soccer, Stefan, Tennis}), containing a diverse range of motion characteristics. To apply the block-wise stationary model, we divided each residual frame into $16 \times 16$ non-overlapping blocks ({\em spatially stationary} blocks) and each such block was divided into $4\times 4$-pixel transform coding blocks (16-dimensional vectors). We then considered a set of time-aligned spatially stationary blocks in 8 adjacent frames to be a locally stationary block of 16-dimensional vectors, and estimated a single covariance matrix for each such {\em spatio-temporal} stationary block. The total training set thus consisted of $26639$ local covariance matrices, which was used as the input to the transform matrix codebook design algorithm.  Various codebook designs were tested using a separate set of 7 video sequences (See Table \ref{tab:VS_BD}.) 

\begin{table}[!tb]
\centering
\caption{\label{tab:Toy_ex}SNR and entropy (bits/sample)  for coding a 3-component Gaussian mixture.}
\begin{tabular}{|c||c|c|}
  \hline
  Transform & SNR (dB) & Entropy\\
      \hline\hline
      KLT & $3.21$ & 0.71  \\
      DCT & $3.69$ & 0.63  \\
      High-rate & $4.0$ & 0.59  \\
      Laplace & $4.0$ & 0.59 \\
      \hline
\end{tabular}
\end{table}

In our experiments we used the same step-size $\Delta$ for all transform coefficient in a vector and set the dead zone to $\left( -\frac{\Delta}{2}, +\frac{\Delta}{2} \right)$. In general, $\Delta$ determines the bit rate. However, when the Laplacian MSE model is used, $\Delta$ required to achieve a given bit rate has to be used in the transform codebook design algorithm, see \eqn{Laplacian_model}. This can be achieved by designing transform codebooks for many $\Delta$ values and selecting the $\Delta$ and the transform codebook that has (approximately) the desired rate. We however found that, a transform codebook designed for an appropriately chosen value of $\Delta$ remains nearly optimal for a wide range of $\Delta$ values. We further elaborate on this point below. Since the DCT will likely be good for some stationary blocks, we included the DCT as an additional codeword, after designing a codebook. In the following, we have used the estimated entropy of the quantization indices as a proxy for the rate of variable-length coding. The total rates reported for adaptive coding include the rates of both the quantized transform coefficients and the codebook indices identifying the transform matrix for each stationary block.

\begin{figure}[!tb]
  \centering
  \includegraphics[width=0.9\textwidth]{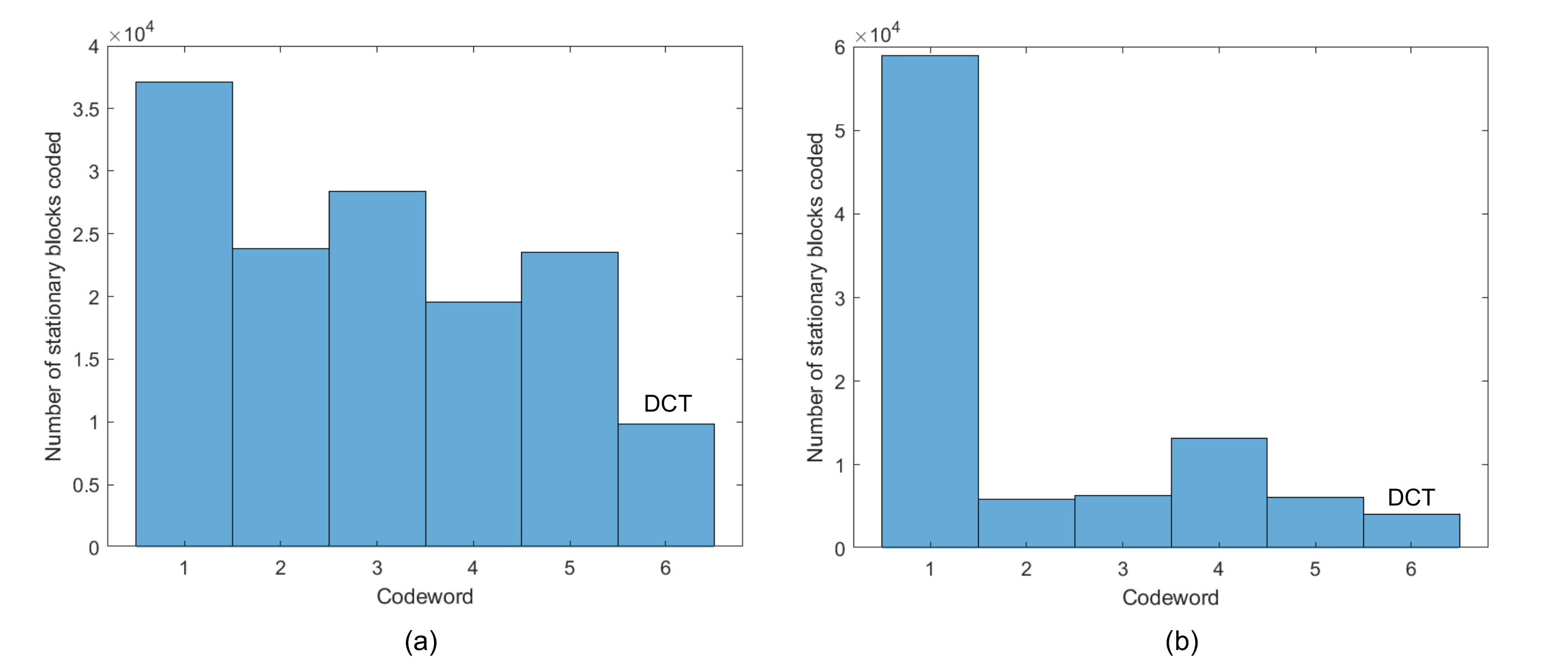}
  \caption{\label{histo} Histograms of transform-matrix codeword usage in (a) Football and (b) Ice sequences. The size of the codebook $N=6$. The first 5 codewords have been optimized using the proposed algorithm, using the Laplacian MSE model(codeword 6 is the standard DCT.)} 
\end{figure}

Table \ref{tab:VS_BD}, presents the gains in peak signal-to-noise ratio (PSNR) and rate achieved by optimized codebooks relative to the DCT, expressed in terms of Bj\o{}ntengaard-Delta (BD) PSNR and BD-bit-rate \cite{Bjontegaard}. Note that, a positive BD-PSNR indicates a gain in PSNR compared to the DCT while a negative BD-rate indicates a rate saving. To further illustrate the advantage of using adaptive transforms over the DCT, we show in Fig. \ref{histo}, the histograms of codeword usage in Football and Ice sequences.
\begin{table}[!tb]
\centering
\caption{\label{tab:VS_BD} BD-PSNR and BD-Rate gains achieved by transform matrix codebooks over the standard DCT. {\em High-Rate} and {\em Laplace} refer to quantization MSE model used for codebook optimization.}
\vspace{0.5ex}

\resizebox{\textwidth}{!}{
\begin{tabular}{@{}l|cccc|cccc|cccc@{}}
 \hline
\multicolumn{1}{c}{\multirow{4}{*}{Sequence}} & \multicolumn{12}{|c}{Codebook size $N$}\\
\cline{2-13} 
\multicolumn{1}{c}{}	& \multicolumn{4}{|c}{3}		& \multicolumn{4}{|c}{6}	 & \multicolumn{4}{|c}{9}\\
\cline{2-13}
\multicolumn{1}{c|}{}	 & \multicolumn{2}{c}{BD-PSNR (dB)} & \multicolumn{2}{c|}{BD-Rate (\%)} & \multicolumn{2}{c}{BD-PSNR (dB)} & \multicolumn{2}{c|}{BD-Rate (\%)} & \multicolumn{2}{c}{BD-PSNR (dB)} & \multicolumn{2}{c}{BD-Rate (\%)} \\
\cline{2-13} 
\multicolumn{1}{c|}{}	& High-Rate        & Laplace       & High-Rate        & Laplace       & High-Rate        & Laplace       & High-Rate        & Laplace       & High-Rate        & Laplace       & High-Rate       & Laplace\\
\cline{1-13}
Akiyo           & 0.096           & 0.148        & -2.86          & -5.10       & 0.122           & 0.185        & -3.60          & -5.39       & 0.135           & 0.192        & -3.94          & -5.43       \\
City            & 0.096           & 0.233        & -1.44          & -3.80       & 0.180           & 0.332        & -2.75          & -5.14       & 0.180           & 0.343        & -2.74          & -5.35       \\
Flower          & -0.009          & 0.152        & 0.15           & -1.44      & 0.048           & 0.221        & -0.41          & -2.04       & 0.048           & 0.211        & -0.39          & -1.88       \\
Hall Monitor    & 0.137           & 0.250        & -3.44          & -6.49       & 0.187           & 0.313       & -4.46          & -7.46       & 0.185           & 0.323        & -4.46          & -7.65       \\
Ice             & 0.157           & 0.393        & -2.91          & -7.31       & 0.180           & 0.467        & -3.27          & -8.27       & 0.179           & 0.465        & -3.29          & -8.28       \\
Mother daughter & 0.224           & 0.290        & -5.83          & -8.66       & 0.259           & 0.367        & -6.61          & -9.08       & 0.264           & 0.379        & -6.74          & -9.58       \\
Waterfall       & 0.152           & 0.239        & -1.86          & -3.53       & 0.281           & 0.356        & -4.10          & -5.00       & 0.292           & 0.354        & -4.30          & -5.17  \\
\hline\hline%
\cline{1-13}                                                                                                                                      
\textbf{Average}	& 0.122	& 0.244	& -2.60	& -5.19	& 0.180	& 0.320	& -3.60	& -6.05	& 0.183	& 0.324	& -3.69	& -6.19
\\
\hline
\end{tabular}}
\end{table}

\begin{figure*}[!tb]
  \centering
 \includegraphics[width=\textwidth]{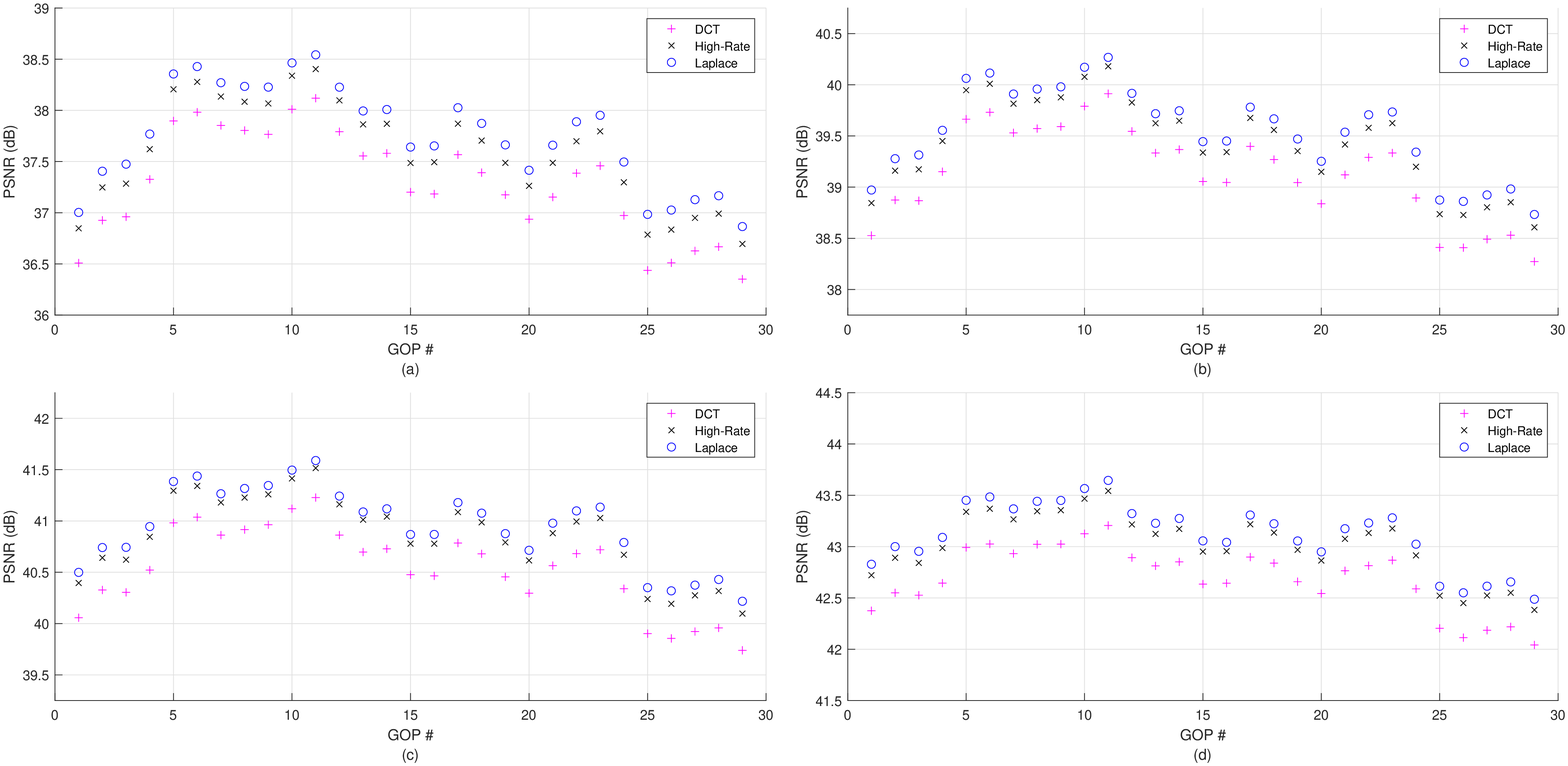}
  \caption{PSNR of coding the Ice sequence using adaptive transforms (codebook size $N=6$) and the DCT (non-adaptive). (a) 0.45 bits/pixel, (b) 0.75 bits/pixel, (c) 1.02 bits/pixel and, (d) 1.44 bits/pixel. PSNR has been computed for groups of 8 consecutive frames.}
  \label{fig_ice}
\end{figure*}

As an example of how adaptive transform coding and non-adaptive DCT compares over a single sequence, Fig. \ref{fig_ice} shows the absolute PSNR (not BD-PSNR) of the MC residual for the Ice sequence, where to avoid clutter in the figure, we compute and show the average PSNR for groups of 8 consecutive frames. As one would expect, the difference between the codebook optimized with the high-rate Gaussian model and the finite-rate Laplacian model diminishes as the rate increases. However, in all our experiments with MC residuals, it was observed that the Laplacian model always yielded a better codebook. This is because at low rates, the high-rate Gaussian model can be quite inaccurate or even outright invalid \cite{goyal2000transform}. Given that the codebooks are designed off-line, the slight complexity increase associated with the use of Laplacian model (compare \eqn{high_rate_model} and \eqn{Laplacian_model}) would be of no consequence. An interesting issue however is the robustness of a codebook designed with the Laplacian model for a specific $\Delta$: how different are codebooks designed for significantly different $\Delta$ values? In HM test model (H265 codec), the parameter QP whose value can range from 0 to 51, is used to set the bit-rate and hence the quantization step-size. Our experiments showed that, a single Laplace codebook optimized for QP=34 is nearly as good as the codebooks optimized for each QP value in the entire range.

\section{Future Work} \label{sec:conclusion}
There exist other descent-type algorithms for optimization on manifolds, which can be explored, see \cite{Manton}. While these algorithms may converge faster, they perform descent steps on geodesics of the manifold which can be computationally quite complex. The algorithm we have used performs descent steps on tangent spaces of the manifold which is much simpler, but results in slower convergence. It seems possible to extend our codebook design procedure to use the Newton-type algorithm for manifold optimization presented in \cite{manton2002optimization}, which also takes descent steps on tangent spaces, but will likely converge faster. Finally, while we have considered only non-separable transforms in this paper, an extension to separable transforms is being currently developed.

\Section{References}
\bibliographystyle{IEEEbib}
\bibliography{IEEEabrv,DCC22_local}

\end{document}